\documentclass[12pt]{article}

\usepackage{sbc-template}
\usepackage{enumerate}
\usepackage{listings}
\usepackage{xspace}
\usepackage{multirow}
\usepackage{booktabs}
\usepackage{graphicx}
\usepackage{subcaption}
\usepackage{stfloats}
\usepackage[T1]{fontenc}
\usepackage{etoolbox}
\usepackage[table]{xcolor}
\usepackage{hyperref}
\usepackage{wrapfig}

\newcommand{\lstfontfamily}{\small\ttfamily}

\definecolor{darkviolet}{rgb}{0.5,0,0.4}
\definecolor{darkgreen}{rgb}{0,0.4,0.2} 
\definecolor{darkblue}{rgb}{0.1,0.1,0.9}
\definecolor{darkgrey}{rgb}{0.5,0.5,0.5}
\definecolor{lightblue}{rgb}{0.4,0.4,1}

\definecolor{stringColor}{rgb}{0.16,0.00,1.00}
\definecolor{annotationColor}{rgb}{0.39,0.39,0.39}
\definecolor{keywordColor}{rgb}{0.50,0.00,0.33}
\definecolor{commentColor}{rgb}{0.25,0.50,0.37}
\definecolor{javadocColor}{rgb}{0.25,0.37,0.75}
\definecolor{jTagColor}{rgb}{0.50,0.62,0.75}
\definecolor{eTagColor}{rgb}{0.50,0.62,0.75}
\definecolor{lineNumberColor}{rgb}{0.47,0.47,0.47}

\definecolor{aliceblue}{rgb}{0.94,0.97,1.0}
\definecolor{beige}{rgb}{0.96,0.96,0.86}

\lstdefinelanguage{Solidity}{
	keywords=[1]{anonymous, assembly, assert, balance, break, case, catch, class, constant, continue, constructor, contract, debugger, default, delegatecall, delete, do, else, emit, event, experimental, export, external, false, finally, for, function, gas, if, implements, import, in, indexed, instanceof, interface, internal, is, length, library, log0, log1, log2, log3, log4, memory, modifier, new, payable, pragma, private, protected, public, pure, push, return, returns, revert, selfdestruct, send, solidity, storage, struct, suicide, super, switch, then, this, throw, transfer, true, try, typeof, using, view, while, with, addmod, ecrecover, keccak256, mulmod, ripemd160, sha256, sha3} 
}

\lstdefinestyle{eclipse}{
  basicstyle=\lstfontfamily,
  emphstyle=\bfseries,
  keywordstyle=\color{keywordColor}\bfseries,
  commentstyle=\markupComments,
  stringstyle=\color{stringColor},
  morecomment=[s][\markupJavadocs]{/**}{*/}, 
  showstringspaces=false,
}

 \graphicspath{ {./imgs/} }

\newcommand{\text}[1]{#1}
\newcommand{\df}{DogeFuzz\xspace}
\newcommand{\dfg}{DogeFuzz-G\xspace}
\newcommand{\dfb}{DogeFuzz-B\xspace}
\newcommand{\dfdg}{DogeFuzz-DG\xspace}
\newcommand{\cf}{ContractFuzzer\xspace}
\newcommand{\sz}{sFuzz\xspace}
\newcommand{\st}{Smartian\xspace}
\newcommand{\iz}{ILF\xspace}
\newcommand{\sds}{\textsc{Bench72}\xspace}
\newcommand{\cds}{\textsc{Bench500}\xspace}
\newcommand{\gb}{grey-box\xspace}
\newcommand{\bb}{black-box\xspace}

\newcommand{\fscore}{$F_1$ score\xspace}

\newcommand{\rqb}{How does the \df bug-finding effectiveness compare to state-of-the-art fuzzers for Ethereum?}
\newcommand{\rqc}{How efficient is \df in fuzzing large-scale, real-world Ethereum smart contracts?}

\title{\df: A Simple Yet Efficient Grey-box Fuzzer for Ethereum Smart Contracts}

\author{Ismael Medeiros\inst{1}, Fausto Carvalho\inst{1}, Alexandre Ferreira\inst{1},\\ 
	Rodrigo Bonif\'{a}cio\inst{1}, Fabiano Cavalcanti Fernandes\inst{2}}
\address{Computer Science Department, University of Brasilia (UnB), Brasilia, Brazil\vspace{-0.6em}
\nextinstitute Federal Institute of Brasilia (IFB), Brasilia, Brazil
\email{\{ismael.medeiros96,faustocarva,alexandre.ps1123\}@gmail.com,\vspace{-0.9em}}
\email{rbonifacio@unb.br,fabiano.fernandes@ifb.edu.br} }

\begin{document}

\maketitle

\begin{abstract}
  Ethereum is a distributed, peer-to-peer blockchain infrastructure that has attracted billions of dollars.
  Perhaps due to its success, Ethereum has become a target for various kinds of attacks, motivating
  researchers to explore different techniques to identify vulnerabilities in EVM bytecode
  (the language of the Ethereum Virtual Machine)---including formal verification, symbolic execution, and fuzz testing.
  Although recent studies empirically compare smart contract fuzzers, there
 is a lack of literature investigating how simpler \gb fuzzers compare to more advanced ones.
  To fill this gap, in this paper, we 
  present \df, an extensible infrastructure for fuzzing
  Ethereum smart contracts, currently supporting \bb fuzzing and two \gb fuzzing strategies: coverage-guided
  \gb fuzzing (\df-G) and directed \gb fuzzing (\df-DG). We conduct a series of
  experiments using benchmarks already available in the literature
  and compare the \df strategies with state-of-the-art fuzzers
  for smart contracts. Surprisingly, although
  \df does not leverage advanced techniques for improving
  input generation (such as symbolic execution or machine learning),
  \df outperforms \sz and \iz, two state-of-the-art fuzzers.
  Nonetheless, the \st fuzzer shows higher code coverage
  and bug-finding capabilities than \df. 
\end{abstract}

\section{Introduction}
\label{sec:INTRODUCTION}


The Ethereum platform was designed to provide a blockchain network enriched with a Turing-complete language that
executes on top of a stack-based virtual machine. Ethereum has gained increased popularity, and developers use different programming languages to implement Ethereum smart contracts---small programs that benefit from a blockchain architecture to ease the implementation of distributed, decentralized, and consensus-based applications. Although developers can use other languages,
Solidity is the most widely used language for implementing Ethereum smart contracts. While expected to support financial digital assets, Solidity and the Ethereum Virtual Machine (EVM) have several design flaws that have allowed bad-intentioned actors to attack the platform, resulting in significant financial losses.

For instance, the infamous DAO attack embezzled four million Ethereum coins (Ethers)~\cite{atzei2017survey} by exploiting a vulnerability of a decentralized autonomous organization implemented as a smart contract. This incident ultimately forced the Ethereum community to implement a hard fork of the entire Ethereum blockchain. The popular Parity Wallet was also a 
target of two attacks that led to significant losses~\cite{palladino2017parity}. 
In a more recent incident, attackers exploited Uniswap's liquidity pool using flash loans to manipulate imBTC coin prices~\cite{openzeppelin_blog}. By inducing a reentrancy attack, they drained a significant number of cryptocurrencies,
showing that smart contracts remain vulnerable to well-known
vulnerabilities such as reentrancy and unhandled exceptions, reflecting the need for more robust
automated security tools.
Common smart contract vulnerabilities (such as reentrancy) are recorded using the Smart Contract
Weakness Classification (SWC).\footnote{https://swcregistry.io/} 
The non-negligible financial impact of attacks targeting the Ethereum ecosystem, coupled with the characteristics of existing smart contracts, has prompted both practitioners and researchers to explore a wide range of techniques for identifying vulnerabilities in smart contracts.

These techniques include static analysis, symbolic execution, formal verification, and fuzz testing (fuzzing)---\cite{survey-attacks} present a survey of Ethereum smart contract vulnerabilities and the techniques used for vulnerability detection in smart contracts.
\cf~\cite{jiang_contractfuzzer_2018} was the first fuzzer designed specifically for smart contracts. Following its initial success, several other research groups began exploring optimization techniques and additional features to enhance the bug-finding capabilities of Ethereum fuzzers.  In this context, we use the term ``bug'' to refer to issues that can eventually introduce a smart contract vulnerability. Consequently, recent fuzzers assert that \cf is outdated and no longer incorporates it into their empirical studies~\cite{nguyen2020sfuzz,he2019learning,harvey}. Advanced fuzzers targeting the Ethereum platform
integrate different strategies for improving their performance, including symbolic execution (e.g., \sz~\cite{nguyen2020sfuzz}),
machine learning (e.g., \iz~\cite{he2019learning}), and a combination of static and dynamic data-flow
analysis (e.g., \st~\cite{choi2021smartian}). Section~\ref{sec:background} presents an overview
of smart contract vulnerabilities in Ethereum and fuzzing for smart contracts.

Although recent studies empirically compare smart contract fuzzers~\cite{wu-fuzzing-icse-2024}, there
is a lack of literature investigating how simpler grey-box fuzzers compare to more advanced ones.
To fill this gap, we present and evaluate \df in this paper. \df is an extensible \gb
fuzzer that currently supports two strategies for prioritizing fuzzer inputs. The first strategy, \df-G, is a
conventional grey-box fuzzer that prioritizes inputs likely to increase code coverage. The second strategy, \df-DG,
prioritizes input more likely to cover instructions close to the so-called critical instructions~\cite{teether}. We present some design decisions of \df in Section~\ref{sec:DESIGN}.
We evaluate \df using a quasi-replication of a previous study~\cite{choi2021smartian} and two benchmarks that we
detail in Section~\ref{sec:SETTINGS}. The results of our empirical assessment (Section~\ref{sec:RESULTS}) revealed
several findings, some of which were unexpected:

\begin{itemize}
\item Although the algorithms \df-G and \df-DG use quite different fuzzing strategies,
  their performance is quite similar. In the larger dataset we use in our experiment,
  both algorithms perform similarly when considering instruction coverage and
  bug-finding capabilities. In the smaller dataset, while \df-G led to higher
  coverage than \df-DG, their bug-finding capabilities are almost identical. 
  
\item Although \df implements simpler fuzzing strategies, \df-G and
  \df-DG outperform state-of-the-art fuzzers (\sz and \iz). Even
  the \bb strategy we implemented achieved a performance
  on par with \sz and \iz. Also, our study
  confirms that the \st outperforms the other fuzzers
  we experiment with (including \df).
\end{itemize}

Altogether, the contribution of this paper is twofold. First, we present the design and implementation of \df,
an open-source, extensible infrastructure for experimenting with fuzzers for smart contracts. Second, we provide an in-depth assessment of \df, demonstrating that simple \gb strategies might outperform \sz and \iz---smart contract fuzzers that rely on state-of-the-art features.

\section{Background and Related Work}\label{sec:background}

\subsection{Smart Contracts Vulnerabilities}

There are several Ethereum vulnerabilities registered
in the Smart Contract
Weakness Classification (SWC) repository.
\cite{survey-attacks} and \cite{atzei2017survey} survey the
field of smart contract vulnerabilities, serving as a good starting point
to understand Ethereum vulnerabilities. 
In this section, we detail a few vulnerabilities that we analyzed and explored
during our research, including Reentrancy, Dangerous Delegate Call, Gasless Send \&
Exception Disorder, and Number Dependency \& Timestamp Dependency. 

{\textbf{Reentrancy (SWC-107)}}  is a design issue that arises from how the
Ethereum Virtual Machine (EVM) handles communication between contracts.
In Ethereum, smart contracts can be invoked using \texttt{CALL} instructions and can define
a \texttt{receive()} method to handle incoming Ether (the native Ethereum cryptocurrency).
Consequently, contracts can include some logic to manage incoming transactions.
The vulnerability known as reentrancy occurs when a method can be repeatedly
invoked within the same transaction by a malicious contract exploiting the \texttt{receive()} method.
This vulnerability arises due to the misuse of \texttt{CALL} instructions without adequately updating
the contract's internal state.

Listing~\ref{code:reentrancy} shows a simple contract that presents this vulnerability.
In this contract, the method \texttt{withdraw()} sends the amount of ether to the caller before it
updates its internal state (\texttt{balances} field). As the method \texttt{call} can be directed to
another smart contract and the EVM resolves the calls sequentially, the method \texttt{withdraw()}
can be called multiple times within a transaction by a malicious contract without the assignment
\texttt{balances[msg.sender] = 0} being executed, draining all the \emph{ether} from that
vulnerable contract.  

\begin{small}
\begin{lstlisting}[float,language = Solidity, caption={A contract with the Reentrancy vulnerability}, label={code:reentrancy}]
contract Reentrancy {
   mapping (address => uint) private balances;
   function withdraw() public {
       uint amount = balances[msg.sender];
       (bool success, _) = 
           msg.sender.call.value(amount)("");
       require(success);
       balances[msg.sender] = 0;
   }
}
\end{lstlisting}
\end{small}

\textbf{Dangerous Delegate Call (SWC-112)} is caused by the use of a variant of the \texttt{CALL}
instruction named \texttt{DELEGATECALL}. This method can be used to execute operations in other
contracts using the context of the caller contract (e.g., the caller contract's balance and variables).
This instruction should be used cautiously to avoid inadvertently calling malicious contracts.
Listing~\ref{code:delegate_call} shows a contract with this vulnerability.
In this case, the method \texttt{fwd()} can be called with any address as argument.
As such, an attacker can pass a malicious contract that will be executed in the contract context (via \texttt{DELEGATECALL}).
This makes the variable \texttt{owner} exposed to be changed by a malicious contract.

\begin{lstlisting}[float,language = Solidity, caption = {A contract with the Dangerous Delegate Call vulnerability}, label=code:delegate_call]
contract DelegateCall {
 address owner;
 constructor() public { owner = msg.sender; }
 function fwd(address c, bytes data) public {
   require(c.delegatecall(data));
 }
}
\end{lstlisting}

\textbf{Gasless Send \& Exception Disorder (SWC-104)} are vulnerabilities related to how an external call is handled.
Each call returns the result of the execution, and when this execution fails, it will return a value to be
handled by the caller contract. If this call is not handled in any way, it will lead to
unexpected behaviors when a failure occurs. 

%

\textbf{Number Dependency \& Timestamp Dependency (SWC-120)} are related to introducing randomness
into smart contracts. Intuitively, one might consider deriving a seed from the number of blocks
or the time a block was mined to implement random computations within a contract method. However,
relying on these data sources can lead to poor randomness computation.

%
%

\subsection{Fuzzing Smart Contracts}

Fuzz testing (or fuzzing for short) is a well-established dynamic approach for finding bugs and
vulnerabilities~\cite{fuzzingbook2024:Fuzzer}.
The central idea of fuzzing is to implement a program (P1) that (a) generates random inputs for another
program (P2) and (b) executes P2 using those random inputs. Over 30 years ago, \cite{miller-original-paper} conducted the first empirical study on ``classic''
black-box fuzz testing for Unix utilities---even though the research on fuzz testing
is still active today, now focusing on newer technologies like mobile apps and smart contracts.

The first fuzzer designed specifically for Ethereum smart contracts, \cf, was developed by \cite{jiang_contractfuzzer_2018}. \cf extends the official Go Ethereum VM implementation by incorporating seven bug oracles that monitor the execution of smart contracts and
identify Ethereum vulnerabilities, such as Exception Disorder and Reentrancy---the latter being
responsible for the infamous DAO attack. Additionally, \cf features a fuzzer that generates random inputs conforming
to the smart contracts' interfaces, as their Application Binary Interfaces (ABIs) specify. Noteworthy, \cf
conducts fuzzing for each function declared in the smart contracts ABI. \cite{jiang_contractfuzzer_2018}
mined smart contracts from a public repository (Etherscan) and assessed \cf's performance
after deploying 6991 smart contracts onto their testnet. In their evaluation, \cf accurately
detected 459 vulnerabilities within these smart contracts~\cite{jiang_contractfuzzer_2018}.

Since then, fuzzing for smart contracts has attracted the interest of different research groups,
and other fuzzers for smart contracts have been developed. Some of the open-sourced fuzzers for smart contracts
include \sz~\cite{nguyen2020sfuzz}, \iz Fuzzer~\cite{he2019learning}, and \st~\cite{choi2021smartian}. \sz is an extensible, feedback-guided fuzzing engine for smart contracts.
Like AFL (a well-known, industry-strength fuzzer for C programs)~\cite{afl}, \sz models the \emph{test generation problem} as an \emph{optimization problem}, prioritizing inputs that cover additional statements in the contract.
Unlike AFL, \sz also prioritizes inputs considering how far a seed is from covering missed
branches~\cite{nguyen2020sfuzz}. \cite{nguyen2020sfuzz} present a comparison between \sz and two other tools: \cf and Oyente~\cite{oyente} (a symbolic execution tool). The
authors report that \sz outperforms the other tools w.r.t the number of test cases generated within a given time interval.
They highlight that since \cf uses a standard EVM implementation, some computational effort is expended while running \cf to
obtain network consensus and to mine blocks. In contrast, \sz executes over an EVM that only simulates network operations relevant to
identifying vulnerabilities in smart contracts. Their empirical evaluation also provides evidence that \sz outperforms
\cf and Oyente regarding branch coverage and the total number of revealed vulnerabilities~\cite{nguyen2020sfuzz}.

The \iz fuzzer for smart contracts employs an \emph{imitation learning} strategy~\cite{imitation-learning}
to enable the test generation process to learn from symbolic execution outcomes---thus generating
inputs capable of traversing deep program paths~\cite{he2019learning}. \cite{he2019learning}
investigated whether \iz achieves higher coverage and discovers more vulnerabilities compared to other security tools for Ethereum,
including the fuzzers Echidna~\cite{echidna} and \cf~\cite{jiang_contractfuzzer_2018}. Their empirical study results indicate that \iz outperforms the other tools in terms of both code coverage and the number of vulnerabilities detected. Furthermore, the authors provide evidence that the \emph{learning component} is essential for achieving higher performance.
Differently, the design of \st aims to identify sequences of transactions that can critically modify
the shared state of smart contracts~\cite{choi2021smartian}, exploring a particular feature of
smart contracts: sequences of transactions share a persistent state. To achieve this, \st
combines static and dynamic analysis of the EVM code. \cite{choi2021smartian}
empirically evaluate \st using a curated benchmark (\sds) that comprises 72 bug-labeled smart contracts.
The results show that \st outperforms other smart contract fuzzers (such as \sz and \iz)
and symbolic execution tools (such as Oyente and teEther). \cite{choi2021smartian} also explore \st in the
wild, using a dataset (\cds) that contains 500 widely used and complex smart contracts.

Other techniques and fuzzers for detecting vulnerabilities in smart contracts do exist, and \cite{survey-attacks}
present a survey on this subject. In this paper, we introduce our 
smart contract fuzzing infrastructure (\df), which currently supports three strategies for input generation:
\bb fuzzing and \gb fuzzing, utilizing both code coverage feedback and critical instruction-guided feedback.
We present the main design decisions and an empirical assessment of \df in Sections~\ref{sec:DESIGN} and~\ref{sec:SETTINGS},
respectively. Surprisingly, \df outperforms \sz and \iz in \sds, even though its fuzzing strategies seem simpler compared
to those employed by these tools. Similar to \cf, \df uses the Go Ethereum implementation of
the EVM.

\textbf{}\section{\df Design Guidelines}
\label{sec:DESIGN}

This section outlines key decisions behind \df's implementation. \df aims to be a flexible, extensible tool for identifying vulnerabilities in Ethereum smart contracts via fuzzing. It currently supports classical \bb, coverage-guided, and a novel directed grey-box fuzzing strategy targeting critical smart contract instructions.

\subsection{\df Building Blocks}

\df's flexibility comes from its web-based architecture, which separates its components: a customized Go Ethereum Virtual Machine, the Vandal Ethereum static analysis tool, the \df fuzzer, data collection, and report modules. The Go Ethereum~\cite{go-ethereum} implementation was chosen for its stability and strong development reputation in the Ethereum community.

Indeed, advanced fuzzing strategies require contextual data to generate new inputs, necessitating EVM customization for collecting execution data. \df gathers dynamic call graphs, instruction execution, and local state changes. This data is used to (1) determine explored paths in a smart contract and (2) capture key EVM events that may indicate vulnerabilities. We customized Go Ethereum to collect this contextual data, inheriting changes from the \cf project and adding instruction collection during transactions. This data is sent to the fuzzer module via an HTTP server.

Also, the \gb fuzzing strategies of \df rely on code coverage data
obtained in real-time during the fuzzer
execution. For that, \df uses an abstract \emph{call-graph} (CG) representation of a smart contract, which
we compute using the Vandal tool~\cite{brent2018vandal}. Vandal is a static analysis tool tailored for Ethreum
smart contracts security. We use the CG representation to compute path coverage for the \gb strategy
and to compute a distance map to critical instructions in the directed \gb strategy.
We believe that opting for a standard EVM implementation like Go Ethereum enables us to
explore Ethereum vulnerabilities across various platform layers~\cite{atzei2017survey}.
This broader exploration wouldn't be feasible with lightweight EVM implementations,
despite their significant performance improvements in fuzzer input generation.
We want to better understand this tradeoff in the long term.

\subsection{Data Gathering}

\df collects and processes metrics during smart contract fuzzing campaigns to generate new inputs. It uses an event-based processing architecture with Go routines for real-time execution metrics. Smart contract calls require parsing inputs into contract function parameters, achieved using a Go Ethereum library that maps Go types to Solidity types. This library also facilitates communication with an Ethereum node for executing smart contract code. These features allow \df to generate inputs, collect execution metrics, use these metrics during fuzzing campaigns, and identify possible vulnerabilities.


\subsection{Fuzzing Strategies}


\df currently supports three fuzzing strategies: \bb, grey-box, and directed grey-box fuzzing.
While the \bb fuzzing strategy only generates random inputs to a given smart contract,
the \gb fuzzing strategy uses coverage data to rank each generated input by the amount of code being explored.
As such, the \gb strategy uses a control flow graph (CFG) to compute the paths of the smart contract and uses the executed instructions to
match each executed branch. \df uses the coverage data of executed branches to compute the best past
execution. With the inputs of these executions, small mutations are made to generate similar inputs that are
more prone to better explore the smart contract.

The directed \gb fuzzing uses a CFG to compute a separated distance map. This map contains the
distances (according to the control-flow graph) to specific instructions classified as critical~\cite{teether}.
We consider four EVM instructions as critical in this work:
\texttt{CALL} (creates a regular transaction), \texttt{SELFDESTRUCT} (makes a contract unusable),
\texttt{CALLCODE} (calls functions in other contracts using the original contract's context), and 
\texttt{DELEGATECALL} (calls functions in other contracts inside the target contract's context).


  
Those instructions were classified in that way because they are related to the studied vulnerabilities.
With that, the directed \gb fuzzing strategy matches the executed instructions to compute how close the
fuzzer is getting to those specific statements of the smart contract. The \df directed \gb strategy uses
that to rank the best inputs to be used as seeds to the next input generation cycle.
These fuzzing strategies were implemented using a common \emph{power schedule} component~\cite{aflgo},
which is responsible for ordering the inputs according to each fuzzing strategy. 
The other important component is how \df detects the vulnerabilities at runtime,
using the data collected by the customization made in the EVM.



\subsection{Bug Oracles}

In order to spot vulnerabilities during fuzzing campaigns, \df leverages the
same approach as other fuzzers: it collects and analyzes event patterns from the
instrumented EVM. Currently, \df collects nine events: Delegate (D), GaslessSend (GS), SendOp (SO), ExceptionDisorder (ED),
BlockNumber (BN), Timestamp (T), Reentrancy (R), StorageChanged (SC), and EtherTransfer (ET).
One or more event combinations can indicate a vulnerability, which the bug oracles analyze.
While the fuzzer explores possible smart contract paths, interpreting execution data is crucial for thorough security analysis. \df uses event sequences to pattern match and identify vulnerabilities in a smart contract. These bug oracles follow a standard implementation,
taking snapshots of transaction events to compute each combination.

\section{Study Settings} \label{sec:SETTINGS}

The goal of our empirical assessment is threefold: (a) to estimate the improvements of \df
\gb fuzzing strategies in comparison with the \bb fuzzing strategy we borrow from \cf,
(b) to investigate how \df performs compared to state-of-the-art fuzzers for smart contracts,
and (c) to understand how \df performs ``in the wild''---considering popular and non-trivial
smart contracts.  
To this end, we conduct two
experiments that aim to answer the following research questions:

\begin{enumerate}[(RQ1)]
\item \rqb
\item \rqc  
\end{enumerate}

(RQ1) deals with the first and second goals of this empirical assessment, while (RQ2) deals with our
third goal. We use a quasi-replication research method, replicating an empirical
assessment conducted in a previous study~\cite{choi2021smartian}
to address our research questions. Throughout our paper, whenever we refer to the ``original study,'' we are referring
to~\cite{choi2021smartian}. Our research validates and extends the findings of the original study by
evaluating two metrics often used by the fuzzing research community: code coverage and bug detection effectiveness.
We compare the \df strategies with three modern fuzzers: \sz, \iz, and \st.

\textbf{Benchmarks.} We use two distinct benchmarks 
in our experiments: \sds and \cds. \sds consists of 72
smart contracts and was curated by the \st team~\cite{choi2021smartian}, who gathered a subset of samples
from the SmartBug benchmark~\cite{durieux2020empirical}. We used \sds to answer (RQ1).
\sds exhibits different classes of vulnerabilities,
totaling 82 labeled bugs across multiple contracts (a single contract might bear more than one bug). The types
of vulnerabilities in \sds are: Block State Dependency (\textbf{BD}) with 13 instances,
Mishandled Exception (\textbf{ME}) with 50 instances, and Reentrancy (\textbf{RE}) with 19 instances. 
Table~\ref{tab:bugoracles} shows a mapping between \df bug oracles,
SWC weakness IDs and the \sds bug oracle names.  \df distinguishes between Timestamp Dependency and Block Number Dependency, but we merge them into Block State Dependency for compatibility with
\sds. Additionally, \st considers Gasless Send, Dangerous Delegate Call, and Exception Disorder as specific cases of
Mishandled Exception.

\begin{table}[ht]
  	\centering	
    \caption{Bug Oracles mapping.} 
    \label{tab:bugoracles}
    \begin{small}
    \begin{tabular}{lcc}\toprule
       \textbf{\df Bug Oracle} & \textbf{SWC} & \textbf{\sds Taxonomy} \\ \midrule
        \text{Reentrancy}              & SWC-107      & RE \\
        \text{Dangerous Delegate Call} & SWC-112      & ME\\
        \text{Gasless Send}            & SWC-104      & ME \\
        \text{Exception Disorder}      & SWC-104      & ME \\
        \text{Number Dependency}       & SWC-120      & BD \\
        \text{Timestamp Dependency}    & SWC-120      & BD \\ 
    \bottomrule
    \end{tabular}
    \end{small}
\end{table}

To address our second research question (RQ2), we utilize a second benchmark called \cds. This dataset includes 500 widely-used smart contracts from the Ethereum mainnet, each with over 30,000 transactions and no labeled bugs (meaning they have not been explicitly identified or marked as containing known vulnerabilities). Nevertheless, some contracts in this benchmark did not compile in our environment due to missing pragma directives and even syntactic errors.
As a result, we excluded 91 samples from our assessment, resulting in our second
dataset with 409 smart contracts. The \st team was also responsible for curating \cds~\cite{choi2021smartian},
selecting smart contracts from Etherscan~\cite{etherscan}.

\textbf{Baselines.} In our experiments, we conduct a comparative analysis between \df and three advanced open-source fuzzers: \sz, \iz, and \st as the baselines. 
\cf was omitted from our evaluation because we also integrated the \bb strategy of \cf into \df. Our selection of \iz \cite{he2019learning} was motivated by its demonstrated superiority over
existing smart contract fuzzers. Additionally, \iz incorporates modern symbolic execution techniques
in conjunction with imitation learning, enhancing its effectiveness. \sz \cite{nguyen2020sfuzz} was chosen
for its foundation on AFL \cite{afl}, a widely recognized fuzzing tool, and its role as a base
for numerous derivative fuzzers \cite{Effuzz,xfuzz}

\textbf{Experimental Procedures.} We conduct different runs of the first benchmark (\sds) for one hour per contract using each tool.
We repeat each experiment five times to minimize statistical errors and then record the average results.
In evaluating all target tools (\iz, \sz, \st), we leverage the same scripts, programs, and Docker images from
the \st team~\cite{choi2021smartian} artifact repository.\footnote{https://github.com/SoftSec-KAIST/Smartian-Artifact}
To ensure compatibility in the \df fuzzing campaigns, we integrate custom scripts and helpers for comparison purposes.
For the second benchmark (\cds), we conduct a 15-minute run for each contract utilizing all three of our fuzzing strategies
(\bb, \gb, and directed \gb). Since this dataset serves as the basis for evaluating coverage across different fuzzing modes,
we deem one fuzzing campaign adequate. We conduct all our experiments using an Ubuntu 20.04 server featuring
42 Intel Xeon CPUs clocked at 2.2 GHz and 432 GB of RAM. We also benefit from Docker version 25.0 in our experiments,
with docker-compose being used for orchestrating \df and a single Docker image used for running the
external tools.

\section{Results}\label{sec:RESULTS}

In this section, we present the results of our empirical assessment and answer our two research questions.
The outcomes of the experiments and all the scripts we use to analyze the results are available in a replication package at
\href{https://doi.org/10.5281/zenodo.11357326}{zenodo}.

\subsection{RQ1: \df Comparison with Other Fuzzers}

\begin{figure}[htbp]
    \centering
    \begin{subfigure}[b]{0.45\textwidth}
      \centering
      \includegraphics[scale=0.50]{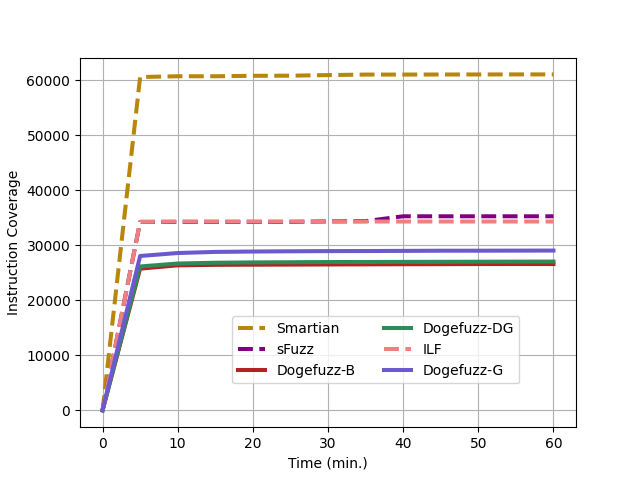}
      \caption{Instruction coverage comparison between \df and other tools on \sds.}
      \label{imgs:sbescovfuzzers}
    \end{subfigure}
    \hfill
    \begin{subfigure}[b]{0.45\textwidth}
      \centering
      \includegraphics[scale=0.50]{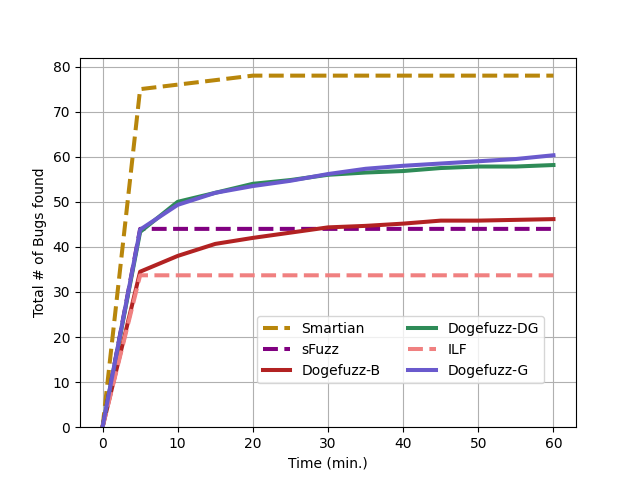}
      \caption{Bug detection comparison against state-of-the-art tools on \sds.}
      \label{imgs:sbesbugsfuzzers}

    \end{subfigure}
    \caption{Comparison of the fuzzers in \sds}
    \label{fig:images}
\end{figure}

To answer our first research question, we compare the different variants of \df with state-of-the-art fuzzers using \sds.
Figures~\ref{imgs:sbescovfuzzers} and~\ref{imgs:sbesbugsfuzzers} summarize the results of the comparison. In the figures, \df \bb fuzzer is denoted as \textbf{\dfb}, the \gb coverage feedback-guided fuzzer is named \textbf{\dfg}, and
the critical instruction directed \gb fuzzer is named \textbf{\dfdg}.
Considering the \df strategies, we can realize in the figure that, initially, all variants
exhibit consistent performance during the first few minutes of testing. Subsequently,
the \dfg variant surpasses the others, primarily due to its strategy of seed prioritization that relies on coverage feedback.

On average, \dfg increases the coverage on this dataset by 9.73\% over \dfb. \dfdg leads to a slight increase in coverage,
showing a 2.2\% improvement over the \bb variant. This suggests that our directed \gb implementation, which prioritizes seeds likely to
execute code close to critical instructions, may also result in achieving higher coverage compared to \bb techniques.
Note that after 30 minutes of execution, the coverage barely increases. This observation is likely due to the \df
lack of more advanced techniques to overcome hard-to-solve conditions (such as using a symbolic
execution tool to enrich the fuzzing strategies).

\df presents the smallest instruction coverage compared to other tools, partly because we filtered out functions with modifiers like view or pure. The reasoning for this decision lies in the fact that these functions, which have no influence on the contract states and are unable to alter state variables, are unlikely targets for exploitation by attackers \cite{li2022redefender}. \df also does not call fallback functions on a non-fallback contract or send ethers to functions that are not marked with the payable keyword. We argue that, despite not achieving higher code coverage compared to other tools, \df successfully altered the persistent state of contracts, leading to the discovery of more bugs
than those identified by \sz and \iz.

So, regarding bug detection comparison,  Figure~\ref{imgs:sbesbugsfuzzers}
illustrates the overall performance of each \df variant. The figure
illustrates the results of our bug-finding comparison over an one-hour period. \dfg and \dfdg
significantly outperform the \bb variant. Interestingly, \dfdg shows an improvement in bug-finding capability of more than 25\% compared
to \dfb, despite only a marginal coverage improvement. \dfg demonstrates a bug-finding capability slightly higher than \dfdg, though.
The results indicate that \dfdg achieves performance levels comparable to \dfg, despite covering fewer instructions.
Although previous studies have examined the application of directed \gb fuzzing to cover recently changed code~\cite{dgbf-ccs},
to the best of our knowledge, we are the first to demonstrate the feasibility of guiding a fuzzer strategy for input
generation that prioritizes seeds that are more likely to execute code close to EVM critical instructions.
In our replication of the original study~\cite{choi2021smartian}, we were able to closely
reproduce the results for \sz and \st. However, regarding \iz, we
encountered an issue within the original study's replication package,
which compromised its ability to detect bugs associated with the Block State Dependency bug oracle.
By addressing this issue, we observed a $\approx$ 32\% improvement in the \iz performance compared to
what was reported in the original study.  We have submitted a pull request with the fix
to the authors of the original study for resolution, but as of now, we have not received a response.

We can see in Figure~\ref{imgs:sbesbugsfuzzers} that \sz and \iz
reach a stable point around 15 minutes, after which no new bugs are found. In contrast, \dfg and \dfdg continue to discover
bugs until the end of the fuzzing campaigns. This is a promising result for \df, assuming that \sz and \iz
are still considered state-of-the-art fuzzers for smart contracts that employ advanced strategies for input generation.
As discussed in the original study~\cite{choi2021smartian}, \st detects substantially
more bugs than \sz and \iz. Figure~\ref{imgs:sbesbugsfuzzers} shows that \st also outperforms the \df strategies,
possibly because it relies on a novel strategy whose fuzzing target is sequences of transactions---instead
of fuzzing individual transactions which is the strategy the other three fuzzers employ.

\begin{wraptable}{l}{0.6\textwidth}
\centering
		\caption{Accuracy by tool for each bug class in Bench72}
		\label{tab:accuracy}
                \begin{tiny}
		\begin{tabular}{p{1.77cm}p{0.42cm}p{0.42cm}p{0.42cm}ccc}
                  
			\hline
			& \textbf{TP} & \textbf{FP} & \textbf{FN} & \textbf{Precision} & \textbf{Recall} & \textbf{\fscore} \\
			\hline
			\multicolumn{7}{c}{\textbf{BlockDependency}} \\
			ILF & 5 & 0 & 8 & 1 & 0.38 & 0.56 \\
			sFuzz & 10 & 0 & 3 & 1 & 0.77 & 0.87 \\
			Smartian & 11 & 0 & 2 & 1 & 0.85 & 0.92\\
			Dogefuzz-G & 9 & 1 & 4 & 0.90 & 0.69 & \cellcolor{gray!25}0.78\\
			Dogefuzz-DG & 9 & 1 & 4 & 0.90 & 0.69 & \cellcolor{gray!25}0.78\\
			Dogefuzz-B & 8 & 0 & 5 & 1 & 0.62 & \cellcolor{gray!25}0.76\\
			\hline
			\multicolumn{7}{c}{\textbf{MishandledException}} \\
			ILF & 11 & 0 & 39 & 1 & 0.22 & 0.36\\
			sFuzz & 29 & 6 & 21 & 0.83 & 0.58 & 0.68\\
			Smartian & 48 & 0 & 2 & 1 & 0.96 & 0.98\\
			Dogefuzz-G & 39 & 9 & 11 & 0.81 & 0.78 & \cellcolor{gray!25}0.80\\
			Dogefuzz-DG & 35 & 7 & 15 & 0.83 & 0.70 & \cellcolor{gray!25}0.76\\
			Dogefuzz-B & 31 & 4 & 19 & 0.89 & 0.62 & \cellcolor{gray!25}0.73\\
			\hline
			\multicolumn{7}{c}{\textbf{Reentrancy}} \\
			ILF & 18 & 2 & 1 & 0.90 & 0.94 & 0.92\\
			sFuzz & 5 & 20 & 14 & 0.20 & 0.26 & 0.26\\
			Smartian & 19 & 0 & 0 & 1 & 1 & 1\\
			Dogefuzz-G & 16 & 4 & 3 & 0.80 & 0.84 & \cellcolor{gray!25}0.82\\
			Dogefuzz-DG & 14 & 4 & 5 & 0.78 & 0.74 & \cellcolor{gray!25}0.76\\
			Dogefuzz-B & 7 & 4 & 12 & 0.64 & 0.37 & \cellcolor{gray!25}0.47\\
			\hline
		\end{tabular}
                \end{tiny}
\end{wraptable}

Table~\ref{tab:accuracy} summarizes our findings for vulnerabilities detected by each tool for each
category of bugs we evaluated. Each column gives the number of true positives (TP), false positives (FP), false
negatives (FN), precision (P), recall (R), and \fscore (F1) for every tool and bug class.
For instance, in the case of Reentrancy (RE), \sz identified 5 true positives out of a total of
19 actual vulnerabilities, with 20 false positives reported.
Overall, \df strategies correctly detect more bugs than \sz ($\approx$ 38,6\%) and \iz ($\approx$ 84,8\%). \dfg and \dfdg
achieve an accuracy (\fscore) above
76\%, contrasting with \sz and \iz that achieve a \fscore close to 60\%. \st leads to higher accuracy, with \fscore of 92\%.
Note that \df did not achieve 100\% precision, even though we reused the same bug oracles as \cf, along with a few fixes we implemented.
\cite{jiang_contractfuzzer_2018} claim that the \cf bug oracles are highly precise and should not generate a
higher number of false positives. However, in our assessment, \df exhibited
lower precision compared to a previous assessment of \cf. We believe that this
difference can be mostly attributed to an improved understanding of the requirements needed to identify a bug that could potentially
lead to a vulnerability, as well as the use of a more accurate dataset that correctly labels the bugs in the smart contracts.

We manually reviewed the contracts labeled with Block Dependency to investigate the
reasons behind the low \df recall rate. During the inspection, we discovered four
contracts that had not been identified as having a Timestamp or Number dependency bug.
These contracts contain conditions that use the Solidity expression \texttt{require} with
specific numbers (magic numbers), which primarily represented ether values or \texttt{if} conditions
with identical arguments. In complex contracts with rigid constraints, it becomes
challenging to satisfy these  ``magic'' numbers. Even fuzzers that incorporate symbolic
execution struggle with these constraints due to path explosion. Other tools in our benchmark
performed better than \df by using advanced strategies to handle this issue.
For instance, \sz uses an adaptive strategy that selects seeds based on a quantitative
measure of how close a test case is to cover any recently missed branch.
Similarly, \st addresses this challenge through concolic testing to determine
the appropriate argument values~\cite{eclipser}.

\subsection{RQ2: \df Assessment in the Wild}

\begin{figure}[htbp]
\centering
\includegraphics[scale=0.40]{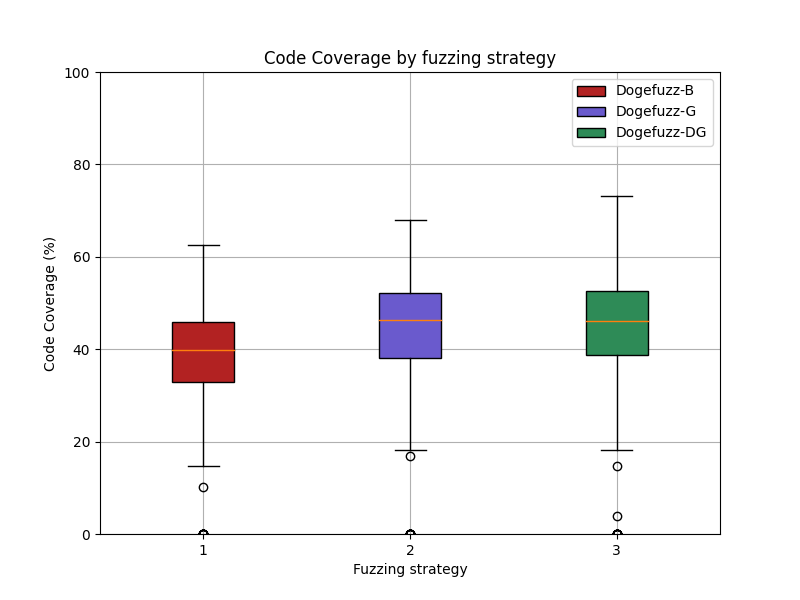}
\caption{Code coverage comparison between \df fuzzing strategies on \cds.}
\label{imgs:sbescovlsfuzzers}
\end{figure}

To answer our second research question, we execute \df variants
\dfb, \dfg, and \dfdg using \cds---which comprises 500 popular Ethereum
smart contracts~\cite{choi2021smartian}. Our goal here is to
understand how \df scales with more complex smart contracts.
Unfortunately, there is missing information in the \cds contracts, preventing \df
from compiling or executing 91 smart contracts. Ultimately, we address RQ2 using 409 contracts from \cds.
Here, we execute each fuzzing campaign for 15 minutes. Figure~\ref{imgs:sbescovlsfuzzers} summarizes the
code coverage achieved using each \df variant. Executing the fuzzers \dfg and \dfdg for 15 minutes led to a
median code coverage of about $\approx$ 48\%, while the black-box strategy achieved a
median code coverage of close to $\approx$ 40\%.

\begin{table}[ht]
  	\centering	
    \caption{Bugs Reported in \cds.} 
    \label{tab:bugs_largescale}
    \begin{small}
    \begin{tabular}{lcc}\toprule
       \textbf{\df Bug}             & \textbf{\df-G} & \textbf{\df-DG} \\ \midrule
        \text{Reentrancy}              & 16      & 13 \\
        \text{Dangerous Delegate Call} & 0      & 0\\
        \text{Gasless Send}            & 29      & 24 \\
        \text{Exception Disorder}      & 2      & 2 \\
        \text{Number Dependency}       & 7      & 8 \\
        \text{Timestamp Dependency}    & 78      & 78 \\ 
        \hline
        Totals & 132 & 125\\
    \bottomrule
    \end{tabular}
    \end{small}
\end{table}

This time, \dfdg achieves a code coverage quite similar to
\dfg. Nonetheless, when considering the bug-finding capabilities,
\dfg found more bugs (see Table~\ref{tab:bugs_largescale}): three
more Reentrancy bugs and five more Gasless Send bugs. Since the
authors of \cds do not label the bugs in each contract,
we cannot investigate the precision and recall of the \df
variants in this experiment. However, we identified that many
of the bugs \dfg found, \dfdg also found. As we mentioned, this is an unexpected result since \dfg
and \dfdg use pretty different strategies for seed prioritization.
A possible explanation for this unexpected result comes
from the fact that smart contracts often contain a small number
of lines of code and conditional statements. As such, both strategies
achieve almost the same set of executed instructions. Additionally,
we found that critical instructions often appear in the contracts
we analyzed (see Table~\ref{tab:ci}), which might actually
cause our guided \gb strategy to yield similar results to the coverage-guided \gb strategy.

\begin{table}[ht]
  \centering
  \caption{Usage of critical instructions in \sds}
  \label{tab:ci}
  \begin{small}
\begin{tabular}{rlrr}
  \toprule
 & Instruction & Total & Average (per contract) \\ \midrule
1 & \texttt{CALL} & 171 & 2.41 \\ 
  2 & \texttt{CALLCODE} &   4 & 1.33 \\ 
  3 & \texttt{DELEGATECALL} &   2 & 1.00 \\ 
  4 & \texttt{SELFDESTRUCT} &  33 & 5.50 \\ 
   \bottomrule
\end{tabular}
\end{small}
\end{table}


\section{Discussion and Threats To Validity}

The results of our empirical study demonstrate the impact of
employing \gb fuzzing strategies in the realm of smart contracts.
Although previous research has already presented the benefits of using
directed \gb~\cite{dgbf-ccs} that use as target recently changed code, our
strategy is novel and, intriguingly, yields a performance almost identical to the classical coverage-guided strategy.
Since we evaluated the \df strategies using only two datasets (already published in the literature),
we cannot generalize our results. Still, we consider our current findings promising,
particularly akin to the design of other fuzzers; the extensibility of \df will
enable us (and hopefully other groups) to experiment with new fuzzing strategies and
their combinations.

The findings of our study have some implications for the field of smart contract fuzzing.
Some of them were unexpected. For instance, our comparative analysis, using well-known accuracy
metrics (i.e., precision, recall, and \fscore),  highlights the efficacy of grey-box fuzzing techniques,
particularly \df-G and \df-DG, performing better than more advanced techniques used by \sz and \iz.
Surprisingly, even employing the \bb strategy of \df, borrowed from \cf, yielded
performance comparable to that of \sz and \iz. This finding is intriguing because
other fuzzers often regard the \cf \bb approach as outdated~\cite{nguyen2020sfuzz,he2019learning,harvey},
resulting in previous research excluding \bb strategies from their empirical studies.
In this paper, we address this gap in the literature. Our research results also provide
evidence that \st performs better than \df. This result might be partially due to
\st fuzzing sequences of transactions, instead of individual transactions that is the
approach \df, \sz, and \iz follows.

As mentioned before, our choice of benchmarks might threaten the validity of
our study. We conducted experiments on two benchmarks derived from \st,
namely \cds and \sds. These benchmarks contain known bugs and consist of large, popular real-world contracts.
Although we evaluated our technique across many contracts, we cannot generalize the results
of this paper to contracts available in other benchmarks.
Another limitation of our study arises from our dependence on Vandal for the
static analysis phase of \df. This tool has not received updates since 2020, and
thus our choice to use Vandal might have slightly compromised the
performance of \df. To address this limitation, future work involves developing or
utilizing an updated version of \df that employs more advanced tools to construct the
call and control flow graphs.
Also, a potential issue arises from the possibility of introducing errors during the
setup and configuration of the empirical study. However, we carefully and systematically
mapped bugs from \st to our bug oracles and tailored scripts and files accordingly.
Indeed, our procedures even revealed a bug in the setup of previous studies, which we
fixed and reviewed the performance measurements of \iz---which is slightly better than
reported in previous studies~\cite{choi2021smartian}. 


\section{Final Remarks}
\label{sec:CONCLUSION}

In this paper, we presented \df, an extensible fuzzer for Ethereum smart contracts that currently supports three strategies: \bb fuzzing, coverage-guided \gb fuzzing, and a novel directed \gb, which aims to prioritize inputs
that approximate the fuzzing campaigns to critical Ethereum VM instructions. We compared the performance of \df using two benchmarks available in the literature (\sds and \cds)~\cite{choi2021smartian}, and our results show that \df \gb strategies present pretty similar results in terms of bug-finding capabilities. Moreover, \df \gb strategies outperform two cutting-edge fuzzers (\sz and \iz) in \sds, even though these fuzzers integrate advanced techniques for input generation (such as symbolic execution and imitation learning). Nonetheless, we found that \st outperforms the \df fuzzing strategies.




\section*{Artefact Availability}

We make our data, scripts and source code publicly available for further investigations in Github at \href{https://github.com/PAMunb/dogefuzz_sbseg_artifact}{dogefuzz\_sbseg\_artifact} and \href{https://github.com/PAMunb/dogefuzz}{dogefuzz}

\bibliographystyle{sbc}
\bibliography{sbseg}

\end{document}